\documentclass[fleqn,usenatbib]{mnras}

\usepackage{newtxtext,newtxmath}

\usepackage[T1]{fontenc}
\usepackage{ae,aecompl}

\def\hi{H{\sc i}}
\def\htwo{H$_{2}$}

\usepackage{graphicx}	
\usepackage{amsmath}	
\usepackage{amssymb}	

\title[Passive disks are not {\rm H{\sc i}}-rich]{xGASS: passive disks do not host unexpectedly large reservoirs of cold atomic hydrogen}

\author[L. Cortese et al.]{
L. Cortese,$^{1,2}$\thanks{E-mail: luca.cortese@uwa.edu.au (LC)}
B. Catinella,$^{1,2}$
R.~H.~W.~Cook,$^{1,2}$
and S. Janowiecki$^{3}$
\\
$^{1}$International Centre for Radio Astronomy Research, The University of Western Australia, 35 Stirling Hw, 6009 Crawley, WA, Australia\\ 
$^{2}$ARC Centre of Excellence for All Sky Astrophysics in 3 Dimensions (ASTRO 3D)\\
$^{3}$University of Texas, Hobby-Eberly Telescope, McDonald Observatory, TX 79734, USA\\
}

\date{Accepted 2020 February 13. Received 2020 February 12; in original form 2019 November 20}

\pubyear{2015}

\begin{document}
\label{firstpage}
\pagerange{\pageref{firstpage}--\pageref{lastpage}}
\maketitle

\begin{abstract}
We use the extended GALEX Arecibo SDSS Survey (xGASS) to quantify the relationship between atomic hydrogen (\hi) reservoir and current star formation rate (SFR) for central disk galaxies. This is primarily motivated by recent claims for the existence, in this sample, of a large population of passive disks harbouring \hi\ reservoirs as large as those observed in main sequence galaxies. Across the stellar mass range 10$^{9}<$M$_{*}$/M$_{\odot}<$10$^{11}$, we practically find no 
passive ($\gtrsim$2$\sigma$ below the star-forming main sequence) disk galaxies with \hi\ reservoirs comparable to those typical of star-forming systems. Even including \hi\ non detections at their upper limits, passive disks typically have $\geq$0.5 dex less \hi\ than their active counterparts. We show that previous claims are due to the use of aperture-corrected SFR estimates from the MPA/JHU SDSS DR7 catalog, which do not provide a fair representation of the global SFR of \hi-rich galaxies with extended star-forming disks. Our findings confirm that the bulk of the passive disk population in the local Universe is \hi-poor. These also imply that the reduction of star formation, even in central disk galaxies, has to be accompanied by a reduction in their \hi\ reservoir.

\end{abstract}




\section{Introduction}
Cold neutral hydrogen is the necessary fuel for star formation. Understanding the physical conditions under which cold gas can collapse and form stars is thus a key step towards building a coherent framework for galaxy formation and evolution. It is now accepted that
molecular hydrogen (\htwo) is more directly connected to star formation on local (i.e., kilo parsec or smaller) scales than the atomic (\hi) gas phase  (e.g., \citealp{bigiel08,leroy08}). 

However, most of the cold gas reservoir in galaxies is in the atomic form (e.g., \citealp{catinella18}, hereafter C18), and the molecular component needs to be continuously replenished by the condensation of \hi\ into H$_{2}$ (i.e., on time-scales typical of the H$_{2}$ depletion time: $\sim$1-3 Gyr; \citealp{saintonge17}). 
Thus, the quenching of the star formation in galaxies can be reached in different ways: by removing/consuming the atomic phase only or both \hi\ and \htwo; by preventing the condensation of atoms into molecules; or by keeping \htwo\ stable against collapse. 

In this paper, we focus on the case of disk galaxies with large \hi\ reservoirs that are not feeding star formation. Examples of gas-rich passive disks do exist \citep{gereb18}, and have been known for decades \citep{pickering97}, but so far only a small number of such objects has been reported. Indeed, while \hi\ content correlates less strongly with star formation rate (SFR) than H$_{2}$, SFR remains one of the global galaxy properties more tightly linked to \hi\ mass, with just a weak secondary dependence on stellar structure \citep{catinella10,brown15,cook19}. This implies that \hi\ mass does not strongly vary with morphology, at fixed SFR, and suggests that passive (i.e., $>$2 $\sigma$ below the star-forming main sequence, SFMS) disks with unexpectedly large \hi\ reservoirs for their SFRs should be rare rather than the norm.

Intriguingly, this conclusion has been recently challenged by \citet[hereafter Z19]{zhang19}, who found that nearly all passive disk galaxies have \hi\ reservoirs as large as those observed in the SFMS. They use data from both the extended GALEX Arecibo SDSS Survey (xGASS, C18) and the Arecibo Legacy Fast ALFA (ALFALFA) survey \citep{alfaalfa05} to show that, for stellar masses 10$^{10.6}<M_{*}/M_{\odot}<10^{11}$, the typical \hi\ reservoir of galaxies does not depend on their position in the M$_{*}$ vs. SFR plane. If confirmed, this may have important  implications on our view of galaxy quenching, as already discussed by \cite{peng19}. However, these results appear in contradiction with the bulk of conclusions so far reached by other analysis of the same xGASS and ALFALFA datasets (e.g., \citealp{catinella10,huang12,brown15,saintonge16,gereb18}; C18). 

Thus, to determine whether or not there is a significant population of \hi-rich passive disk galaxies, here we use xGASS to investigate the variation of the \hi\ content across the stellar mass vs. SFR plane, as a function of both stellar mass and bulge-to-total ratio. This not only allows us to revisit the findings presented in Z19 but, most importantly, to extend such analysis to a wider range of stellar masses and morphologies, thus providing stronger constraints on the abundance of passive disks with unexpectedly large \hi\ reservoirs. 

\section{Data}
We use the xGASS representative sample across the stellar mass range 10$^{9}<M_{*}/M_{\odot}<10^{11}$. As extensively described in C18, this sample should be representative of the \hi\ gas content of galaxies in the local Universe (0.01$<z<$0.05), while also reproducing the slope and scatter of the 
SFMS (e.g., \citealp{saintonge16,janowiecki20}). In order to minimize the potential effect of environmental processes in regulating the \hi\ content of galaxies, this paper focuses only on central systems according to the classification performed by \cite{yang07}, including the minor revisions described in \cite{janowiecki17}. With the term `central' galaxies, we refer to both isolated centrals and group centrals: 667 galaxies in total. We include both \hi\ detections and non detections (which will be plotted at their  5$\sigma$ upper limit). We discard 25 galaxies classified as potentially confused in \hi. 

Stellar masses (M$_{*}$) are extracted from the Max Planck Institute for Astrophysics (MPA)/Johns Hopkins University (JHU) value-added catalog based on the Sloan Digital Sky Survey (SDSS) DR7 \citep{sdssDR7}, and assume a \cite{chabrier} initial mass function. Our SFR estimates are computed primarily by combining near-ultravioled (NUV) from GALEX \citep{martin05} with mid-infrared (MIR) fluxes from the Wide-field Infrared Survey Explorer \citep{wright10}. We refer the reader to \cite{janowiecki17} for a more extensive description of our SFR estimates. As we show in this work, the choice of SFR indicator is critical for this type of analysis. Thus, we also test our results against the MPA/JHU DR7 SFRs, which are based on the technique developed by \cite{brinkman04}. Briefly, fiber-based SFRs are derived from the SDSS spectra using the H$\alpha$ line flux for star-forming galaxies, and an empirical relation between specific SFR and D4000 index for active galactic nuclei or weak emission line galaxies. Aperture corrections are then estimated by deriving out-of-fiber SFRs via spectral energy distribution (SED) fitting to SDSS ugriz photometry. 

Lastly, to isolate disk-dominated galaxies we take advantage of the bulge-to-disk decomposition of xGASS galaxies recently presented by \cite{cook19}. This has been derived from $g$, $r$ and $i$ SDSS imaging data using the Bayesian code \textsc{ProFIT} \citep{robotham17}. Stellar mass bulge-to-total (B/T) ratios are then determined from the $r$-band profile and the $g-i$ color gradients using the empirical recipes presented in \cite{zibetti09}. For 57 centrals in our sample either NUV+MIR SFRs are not available and/or bulge-to-disk decomposition was not feasible (e.g., interacting/disturbed systems). Thus, our final sample consists of 585 central galaxies across the 10$^{9}<M_{*}/M_{\odot}<10^{11}$ mass range: i.e., $\sim$90\% of our not confused sample, thus most likely representative of the local galaxy population. Indeed, out of the 57 galaxies removed, 10 are not detected in \hi\ and 32 have SFRs greater than 10$^{-0.5}$ M$_{\odot}$ yr$^{-1}$ according to both SFR indicators used in this work (i.e., they lie on the SFMS). In other words, we are excluding no more than 2\% of potentially passive, gas-rich systems. 

\section{Are passive disks \hi-rich?}
\begin{figure*}
	\includegraphics[width=17cm]{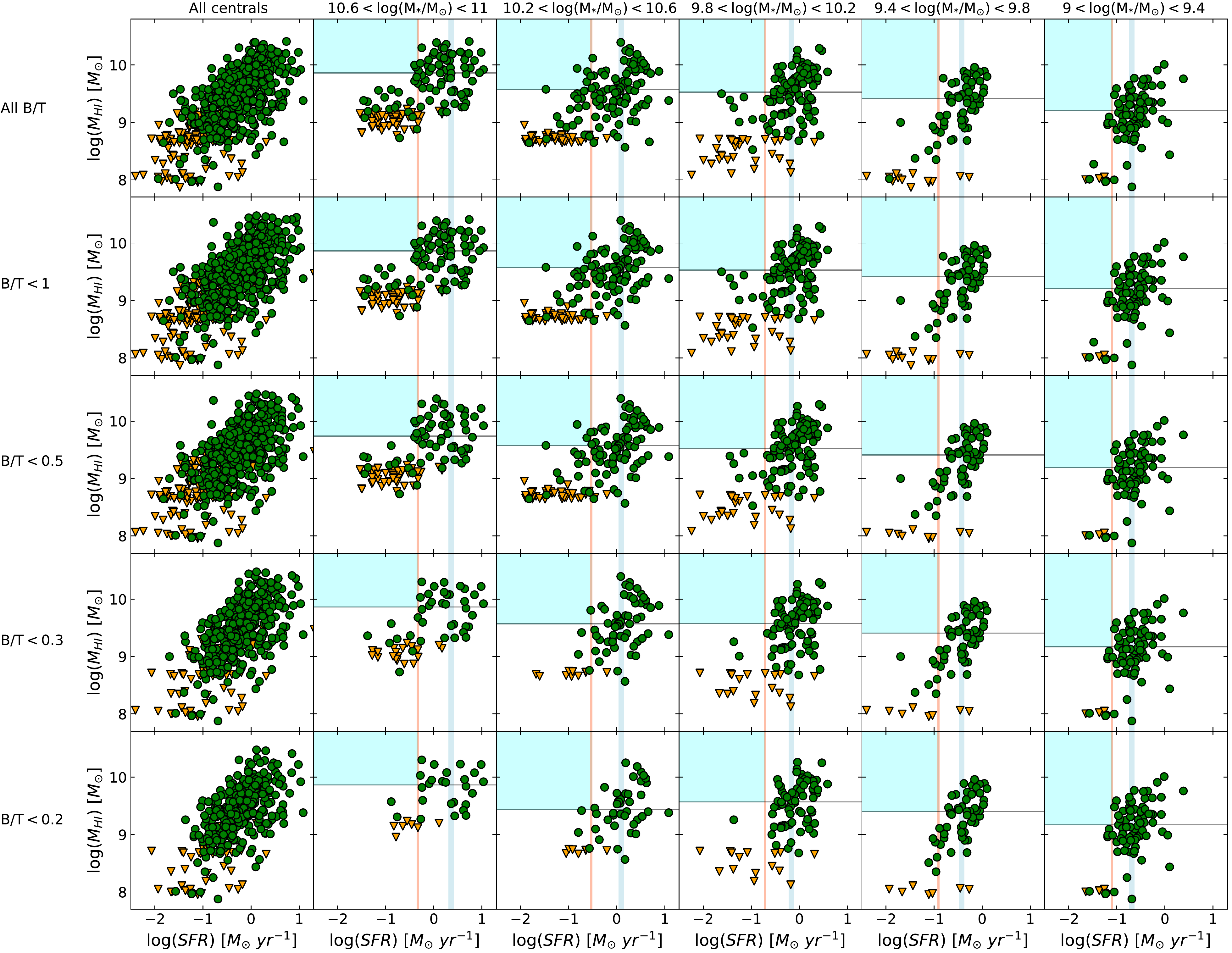}
    \caption{The SFR vs. \hi\ mass relation for the 585 central galaxies in our sample (top left panel), and for different subsamples selected according to stellar mass (columns) and bulge-to-total mass ratio (rows). \hi\ detections and non detections are indicated by green circles and orange triangles, respectively. In each panel, the blue, red and black lines indicate the best-fit SFR for main sequence galaxies, the $2\sigma$ lower limit of the SFMS and the median \hi\ mass of main sequence galaxies in that stellar mass bin, respectively. Passive, \hi-rich galaxies should occupy the top-left corner of each sub-plot, highlighted in cyan.}
    \label{goodsfr}
\end{figure*}

Inspired by the approach adopted by Z19, we look for a large population of \hi-rich passive galaxies by plotting the global \hi\ content of our central galaxies as a function of their SFRs. This is shown in the top left panel of Fig.~\ref{goodsfr}, where \hi\ detections and upper limits are plotted as green circles and orange triangles, respectively. It is clear that \hi\ mass and SFR show a correlation, although with a significant scatter. Interestingly, we find no  population of galaxies in the top-left corner (i.e., passive and \hi-rich). 

However, interpreting this plot is not trivial as we are mixing galaxies of different stellar masses. Thus, the observed correlation could just be a stellar mass effect, due to the fact that lower mass galaxies tend to have both lower \hi\ mass and SFR. Moreover, we are including all galaxies regardless of whether they host a disk or not. Thus, most of the passive population could be dominated by pure bulges, giving the false impression that the two quantities are physically correlated. 

To address these issues, we split galaxies according to both their stellar mass and bulge-to-total mass ratios. In detail, each column of Fig.~\ref{goodsfr} shows a different stellar mass bin (0.4 dex wide), while each row includes a different range of B/T mass ratios, gradually moving from all galaxies at the top towards only pure disks at the bottom. Each panel shows the typical SFR for main sequence galaxies of that stellar mass (blue vertical line), 2$\sigma$ below the SFMS (red vertical line, which we consider the threshold separating star-forming from quiescent systems) and the median \hi\ mass value for main sequence galaxies (black horizontal line). We adopt the fit to the SFMS presented in C18. Although our definition of SFMS is different from the one used in Z19, the threshold chosen to separate star-forming and passive systems is not too dissimilar: i.e., for 10$^{10.6}$<$M_{*}/M_{\odot}$<10$^{11}$ we adopt $\log(SFR~[\rm M_{\odot}~yr^{-1}])\sim-$0.33, whereas Z19 uses $\sim-$0.25 (see their Fig. 1). As expected, for each B/T ratio, lower mass star-forming galaxies have lower amount of \hi\ (e.g., for the top row we find median $\log(M($\hi)$/M_{\odot}))$=9.86, 9.57, 9.53, 9.42, 9.21 moving from high- to low-mass systems), naturally following from the M$_{*}$ vs. \hi\ mass relation \citep{cook19}.

The key point of this figure, and the main result of this paper, is that for every permutation of M$_{*}$ and B/T 
we do not find almost any passive galaxies with \hi\ masses comparable or above the median value observed for the SFMS (i.e., the parameter space highlighted in cyan in each sub-plot). While, above our SFR threshold, \hi\ mass and SFR are loosely correlated (so that galaxies $\sim$1-2 $\sigma$ below and above the locus of the SFMS may have similar \hi\ masses; \citealp{janowiecki20})\footnote{Note that, by construction, the single SFR threshold used to separate star-forming from passive systems encompasses the so-called transition region or green valley (e.g., 1-2$\sigma$ below the SFMS). This is where \cite{janowiecki20} find galaxies that may still host significant gas reservoirs, which is entirely consistent with our results (see their Fig. 5).}, for SFRs more than $\sim$2$\sigma$ below the SFMS the \hi\ reservoir of galaxies dramatically drops and only a couple of objects in the $\sim$10$^{10.4}$ M$_{\odot}$ stellar mass bin are consistent with that observed on the SFMS. Most importantly, even considering non detections at their upper limits, passive galaxies have \hi\ masses $\geq$0.5 dex lower (depending on the stellar mass and B/T bin considered) than what is observed in active galaxies. In other words, {\it a significant population of passive disks, with \hi\ reservoirs as massive as those observed in star-forming systems, does not exist in a sample such as xGASS}.

\begin{figure*}
	\includegraphics[width=17cm]{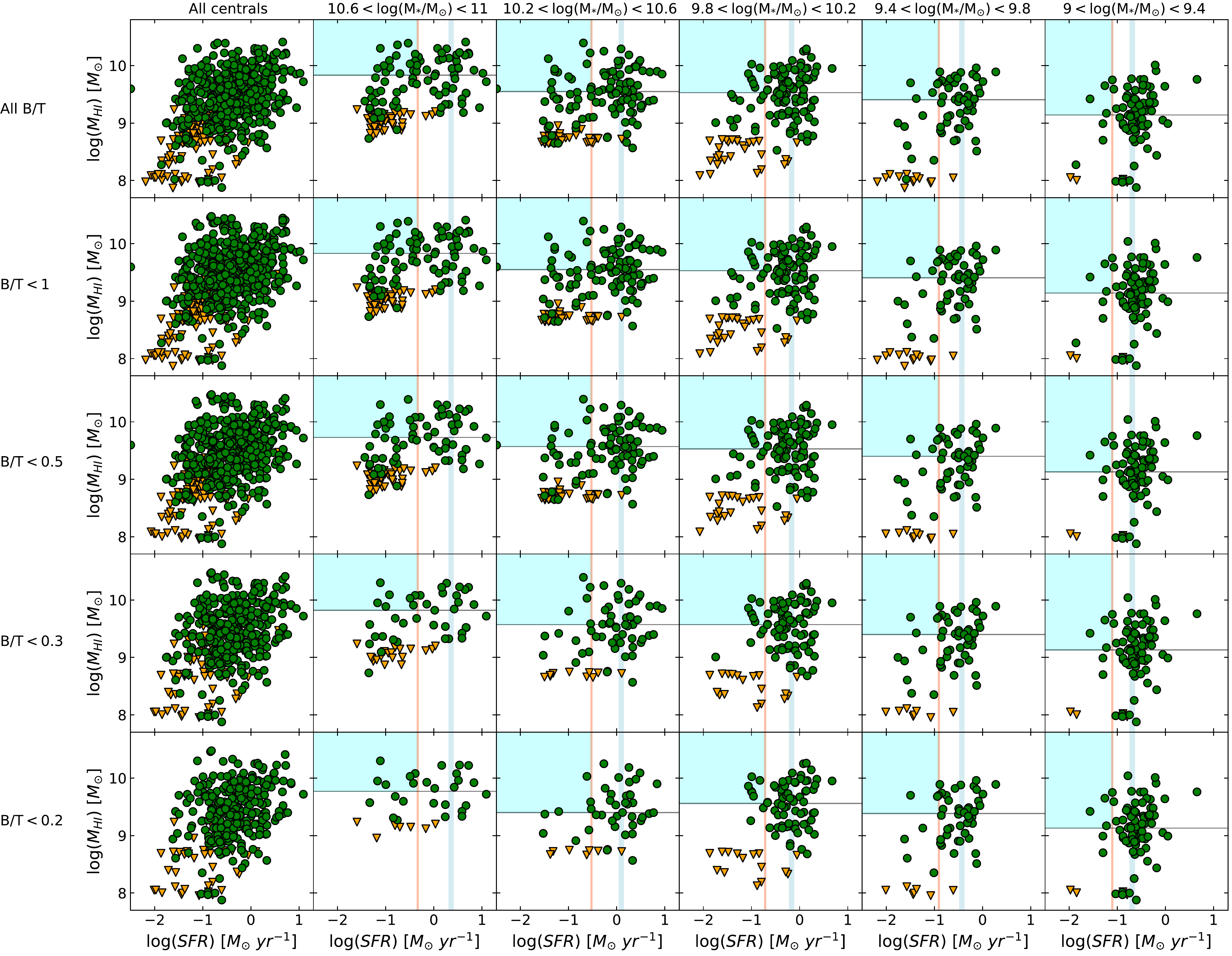}
    \caption{Same as Fig.~\ref{goodsfr}, but using SDSS-based SFRs instead of NUV+MIR SFR estimates.}
    \label{sdsssfr}
\end{figure*}

Our results may appear in contradiction with the recent findings by Z19, who suggest that nearly all passive disks in xGASS are \hi-rich. 
There are two primary differences between the approach used by Z19 vs. ours: the SFRs used (SDSS-based vs.  NUV+MIR) and the selection of disk galaxies (Galaxy Zoo visual classification from \citealp{lintott11} vs. stellar mass B/T ratio). The SFR choice turns out to be the most critical factor in explaining the difference. Indeed, in Fig.~\ref{sdsssfr}, we reproduce Fig.~\ref{goodsfr} using this time the SDSS-based SFRs. 

As can be seen, the difference is dramatic. First, the overall correlation between \hi\ mass and SFR (top left panel) becomes weaker (the Pearson correlation coefficient  drops from $\sim$0.67 to $\sim$0.51). Second, at all stellar masses, and in particular in the stellar mass range 10$^{10.6}<M_{*}/M_{\odot}<10^{11}$ investigated by Z19, we find a non-negligible fraction of gas-rich disks outside the SFMS. The significant difference between these two SFR indicators was already pointed out by Z19 (see their Appendix A), where they compare their results with that obtained using our SFRs estimates as well as those based on SED fitting presented by \cite{salim16}. Indeed, their result clearly shows that, for both NUV+MIR- and SED-based SFRs, no \hi-rich disk galaxies exist below $\log(SFR~[\rm M_{\odot}~yr^{-1}])\sim-$0.4 (i.e., roughly 2$\sigma$ from the SFMS). Such a dramatic change in dynamic range (a factor of $\sim$10) was only partially noticed, leading to the wrong conclusion that both SFR indicators provide the same picture. This is also exacerbated by the fact that the fit adopted by Z19 for the SFMS is not applicable to xGASS in this stellar mass bin for NUV+MIR SFRs, as it does not take into account the bending of the SFMS. Conversely, the difference in the definition of disk galaxies between the two works is less important. Indeed, the Galaxy Zoo disk classification includes galaxies spanning nearly the full range of B/T in our sample. Most importantly, even if we restrict the analysis to pure disks only ($B/T<$0.2), passive central disk galaxies remain an extremely rare population. 

\begin{figure*}
	\includegraphics[width=17.cm]{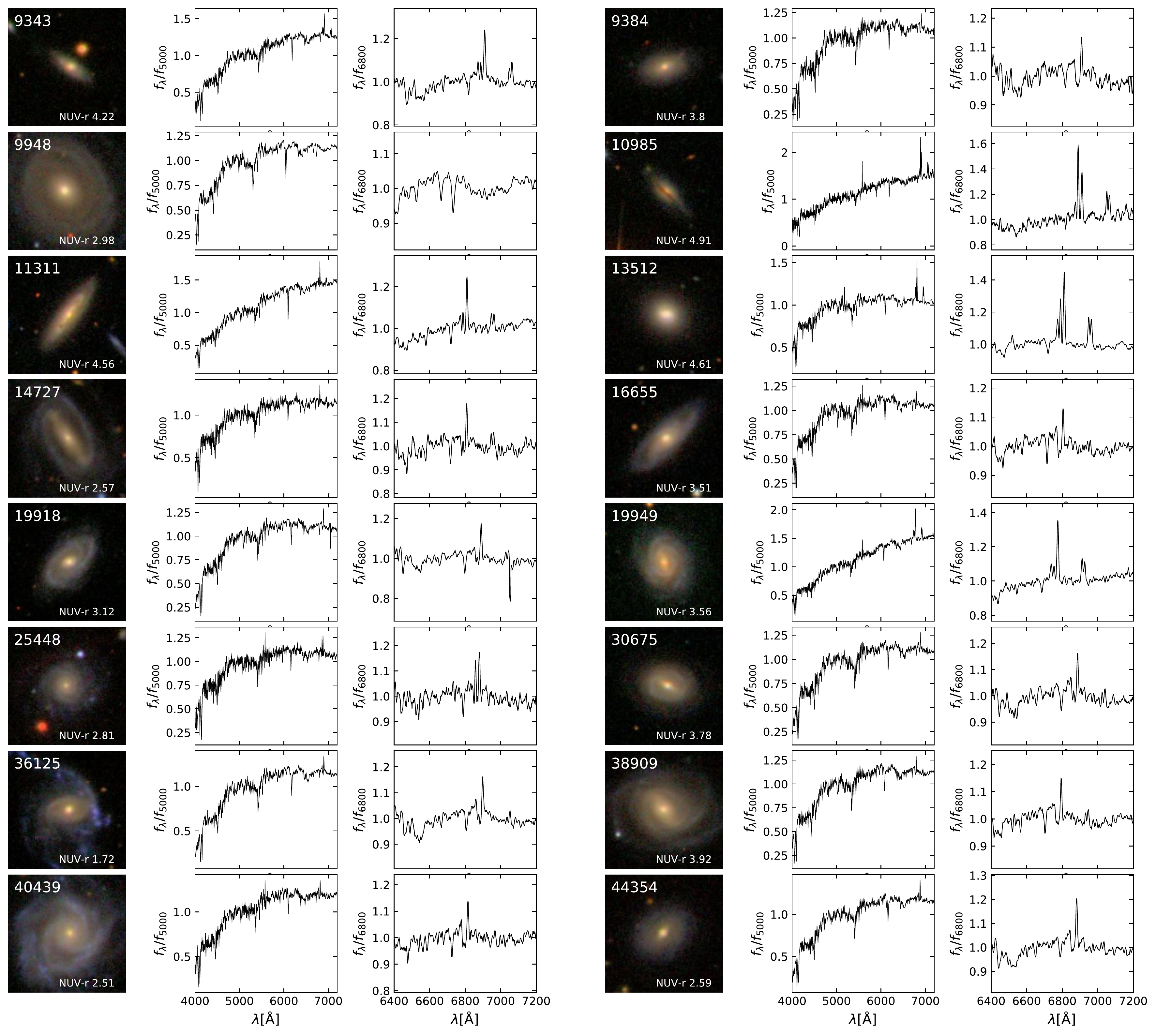}
    \caption{The optical properties of the 16 galaxies in the 10$^{10.6}<M_{*}/M_{\odot}<10^{11}$ stellar mass range which are passive according to the SDSS-based SFRs but have \hi\ mass above the median value of main sequence galaxies. For each galaxy, we show the SDSS colour image (1\arcmin\ size, with xGASS identifier indicated in the top-left and NUV-r color in the bottom right corner, respectively), the optical spectrum and a zoom-in on the spectral region of the H$\alpha$ line.}
    \label{collage}
\end{figure*}

\section{Discussion \& Conclusion}
The results presented here highlight how any conclusions on the existence of a large population of passive \hi-rich galaxies depend on the choice of SFR indicator.
This is most likely due to the known systematic difference between SDSS-based and SFR estimates including UV and infrared data for galaxies whose SDSS-fiber spectrum is either contaminated by AGN emission and/or shows weak emission lines \citep{salim16}.
In order to check if this is the case for xGASS, in Fig.~\ref{collage} we present the SDSS color images and fiber spectra for all the galaxies in the 10$^{10.6}<M_{*}/M_{\odot}<10^{11}$ stellar mass range which are passive according to the SDSS-based SFRs and have total \hi\ masses above the median value for main sequence galaxies. These are the objects that drive the remarkable difference between Fig.~\ref{goodsfr} and \ref{sdsssfr}, and the members of the claimed population of passive \hi-rich disks. 

Even without applying any cut in bulge-to-total ratio, it is clear that practically all these objects (with perhaps just the exception of GASS13512) would be visually classified as disk galaxies. Most importantly, the bulk of this population clearly shows blue spiral arms indicating active star formation. This is quantitatively confirmed by their NUV-r colour, which is consistent with what observed for the SFMS (1.7$<NUV-r~\rm[mag]<$4.5). Thus, these are not passive disks. However, all these objects show SDSS fiber spectra consistent with an old stellar population, with either no emission lines or weak emission lines generally inconsistent with photoionisation by star formation. Indeed, also from the SDSS colour images, it is evident that this family of systems is characterised by an inner (more passive) component and an outer star-forming disk. 

In addition to the visual evidence presented here, the fact that the overall scatter in the SFR-HI mass relation is significantly smaller when UV+MIR-based SFRs are used, and that these are consistent with those independently obtained via spectral energy distribution fitting \citep{janowiecki17}, provide additional support to the fact that Fig.~\ref{goodsfr} gives a more reliable representation of the connection between the integrated \hi\ mass and SFR in nearby galaxies. This is also in line with the fact that \hi\ is more tightly linked to unobscured SFR primarily traced by ultraviolet emission. Thus, any SFR indicator heavily based on the inner parts of galaxies and/or not including ultraviolet information may provide a biased view on the physical link between \hi\ reservoir and star formation activity. Our results also confirm that, at fixed stellar mass and within $\sim$2 $\sigma$ from the locus of the SFMS, \hi\ and SFR are only loosely linked (see also \citealp{janowiecki20}). 

Lastly, Z19 find a significant population of \hi-rich, passive disks also in the ALFALFA survey. Thus, one might wonder whether ALFALFA could include a large population of \hi-rich, passive disks not present in xGASS. This is particularly important given the significantly larger volume covered by ALFALFA with respect to xGASS. However, it is important to clarify that this question has already been answered by Fig. 5 in Z19. Here it is shown that, even for ALFALFA, if SED-based SFR estimates are adopted, no galaxies with stellar mass in the 10$^{10.6}$-10$^{11}$ M$_{\odot}$ range and with \hi\ mass greater than $\sim$10$^{9.5}$ M$_{\odot}$ exist below $\log(SFR~[\rm M_{\odot}~yr^{-1}])\sim-$0.4, i.e., entirely consistent with our findings. In conclusion, we confirm that \hi-rich (i.e., with typical \hi\ masses comparable to those observed in the SFMS) passive (i.e., $\gtrsim$2$\sigma$ below the SMFS) disk galaxies are the exception rather than the norm.

\section*{Acknowledgements}
We thank the referee, Jarle Brinchmann, for useful comments and constructive criticisms which improved the quality of this manuscript. We thank Yingjie Peng and Chengpeng Zhang for useful discussions. This research was supported by the Australian Research Council Centre of Excellence for All Sky Astrophysics in 3 Dimensions (ASTRO 3D), through project number CE170100013. LC is the recipient of an Australian Research Council Future Fellowship (FT180100066) funded by the Australian Government.




\bibliographystyle{mnras}
\bibliography{main} 




\bsp	
\label{lastpage}
\end{document}